\documentclass[conference]{IEEEtran}
\IEEEoverridecommandlockouts
\newcommand{\new}[1]{#1} 
\usepackage{cite}
\usepackage{amsmath,amssymb,amsfonts}
\usepackage{algorithmic}
\usepackage{graphicx}
\usepackage{textcomp}
\usepackage{xcolor}
\usepackage{bm} 

\usepackage{booktabs} 
\usepackage{multirow} 
\usepackage{subcaption} 
\usepackage[bookmarks=false, hidelinks]{hyperref} 
\usepackage{siunitx} 
\newcommand{\vect}[1]{\mathbf{#1}} 
\newcommand{\mat}[1]{\mathbf{#1}} 
\def\j{\mathrm{j}} 
\newcommand{\degree}{\ensuremath{^\circ}} 

\begin{document}

\title{MARBLE-Net: Learning to Localize in Multipath Environment with Adaptive Rainbow Beams
\thanks{This work is supported by the National Natural Science Foundation of China (NSFC) under Grant 62125108 and by the Science and Technology Commission Foundation of Shanghai under Grant 25DP1501200. The source code of MARBLE-Net is publicly available at \url{https://github.com/SJTU-WirelessAI-Lab/MARBLE-Net}.}}

\author{
    \IEEEauthorblockN{Qiushi Liang, Yeyue Cai, Jianhua Mo, and Meixia Tao}
    \IEEEauthorblockA{School of Information Science and Electronic Engineering, Shanghai Jiao Tong University, Shanghai, China\\
    Emails: \{lqs020206, caiyeyue, mjh, mxtao\}@sjtu.edu.cn}
}

\maketitle

\begin{abstract}
Integrated sensing and communication (ISAC) systems demand precise and efficient target localization, a task challenged by rich multipath propagation in complex wireless environments. This paper introduces MARBLE-Net (Multipath-Aware Rainbow Beam Learning Network), a deep learning framework that jointly optimizes the analog beamforming parameters of a frequency-dependent rainbow beam and a neural localization network for high-accuracy position estimation. By treating the phase-shifter (PS) and true-time-delay (TTD) parameters as learnable weights, the system adaptively refines its sensing beam to exploit environment-specific multipath characteristics. A structured multi-stage training strategy is proposed to ensure stable convergence and effective end-to-end optimization. Simulation results show that MARBLE-Net outperforms both a fixed-beam deep learning baseline (RaiNet) and a traditional k-nearest neighbors (k-NN) method, reducing localization error by more than 50\% in a multipath-rich scene. Moreover, the results reveal a nuanced interaction with multipath propagation: while confined uni-directional multipath degrades accuracy, structured and directional multipath can be effectively exploited to achieve performance surpassing even line-of-sight (LoS) conditions.


\end{abstract}


\section{Introduction}
The evolution toward 6G networks features the deep integration of sensing and communication, giving rise to the paradigm of Integrated Sensing and Communication (ISAC) \cite{Gonzalez-Prelcic2025Six}. Leveraging large bandwidths and massive antenna arrays, ISAC enables high-precision localization as a native network service \cite{Jabeen2025Localization}. However, this integration exposes a dichotomy in how wireless systems interact with multipath propagation. For communications, multipath is a valuable resource that enables diversity and multiplexing gains \cite{Alkhateeb2014MIMO}; for monostatic sensing, it is often regarded as interference that causes ambiguities and degrades performance \cite{jia2024multipath, lu2022degrees}.

This challenge has spurred a paradigm shift from multipath mitigation to multipath exploitation, where multipath components are reinterpreted as rich sources of information.
In multipath-rich environments, echoes via non-line-of-sight (NLoS) paths provide additional observation perspectives and spatial diversity, improving sensing accuracy, particularly for targets with fluctuating radar cross-sections (RCS) \cite{zhang2024multipath, lv2025target}.
For extended targets such as vehicles, multipath echoes can even exceed direct reflections and carry more precise motion information \cite{zhao2025sensing}.
The goal is thus to transform multipath “from enemy into friend” by exploiting its embedded geometric cues for localization tasks \cite{liu2025multipath}.

A highly efficient physical-layer mechanism for spatial exploration in wideband ISAC systems is frequency-dependent beamforming, which forms ``rainbow beams'' through controlled beam squint\cite{Yan2019Wideband, luo2023beam, Cui2023NearField, gao2023integrated, Zhou_Gui_JSAC25, Zheng2025Near}. By employing true-time-delay (TTD) devices, often as part of a phase-time array (PTA) \cite{ratnam2022joint, Mo_Jianhua_Asilomar24,cai2025hybrid, Liang2025CFARNet, nam2025joint}, this technique uniquely maps different frequency subcarriers to distinct spatial locations, enabling a wide angular sector to be scanned within a single OFDM symbol. This ``single-scan'' operation reduces overhead and latency, catering to real-time applications. The beam squint effect, usually viewed as a communication impairment, is thus harnessed to encode angular information directly in the frequency domain.

Despite this efficiency, interpreting the complex multipath-rich echoes from rainbow beams remains challenging \cite{mokri2024custom}.
The superposition of numerous propagation paths forms a unique frequency-domain ``fingerprint'' for each target, too intricate for conventional estimators that rely on dominant LoS paths.
This complexity, however, creates an opportunity for data-driven methods.
Deep learning (DL) excels at modeling non-linear mappings from high-dimensional signals to geometric outputs \cite{papageorgiou2021deep, chen2024sdoa}.
While prior works apply DL for DoA estimation \cite{chen2024sdoa} or near-field localization with fixed rainbow beams \cite{klus2025deep}, jointly leveraging multipath propagation and adaptive beam squint within a unified neural framework remains largely unexplored.

Most existing methods, including DL-based ones, rely on fixed, pre-designed rainbow beams that are non-adaptive and often suboptimal for specific environments \cite{klus2025deep}. To overcome this limitation, we adopt a co-design paradigm that jointly optimizes the sensing beam and estimation network \cite{Heng_TWC22}. By treating the beamforming parameters—phase-shifter (PS) and TTD values—as learnable weights in an end-to-end framework, the system autonomously shapes its rainbow beams and exploits environmental multipath information. Building on this idea, we propose a novel deep learning framework for single-target localization that jointly optimizes the parameters of a TTD-based rainbow beamformer and a subsequent convolutional neural network (CNN) for position estimation. Our key contributions are summarized as follows:
\begin{itemize}
    \item \textbf{Synergistic Co-design for Adaptive Sensing:} We propose a fully differentiable, end-to-end model, named \textbf{MARBLE-Net} (\textbf{M}ultipath-\textbf{A}ware \textbf{R}ainbow \textbf{B}eam \textbf{LE}arning \textbf{Net}work), that jointly optimizes a learnable rainbow beamformer and a CNN-based location prediction network. This co-design enables the system to adapt its physical-layer sensing beam to the specific propagation environment via gradient-based optimization.

    \item \textbf{Structured Training Methodology:} \new{We propose a three-stage training strategy that (i) initializes the learnable rainbow beamformer ($\mathcal{M}_1$) by maximizing the total received signal power while keeping the prediction network ($\mathcal{M}_2$) fixed to increase the uplink SNR, (ii) alternates between updating $\mathcal{M}_2$ and adapting $\mathcal{M}_1$ to minimize the localization error, thereby steering the beam pattern to exploit geometrically informative multipath components, and (iii) performs end-to-end joint fine-tuning of $\mathcal{M}_1$ and $\mathcal{M}_2$.}
\end{itemize}




\section{System and Channel Model}
We consider an uplink localization scenario where a base station (BS) aims to determine the 2D position of a single target, modeled as an unmanned aerial vehicle (UAV). The BS is equipped with a uniform linear array (ULA) of $N_r$ antenna elements. Both the BS and the UAV are assumed to be operating at a height of \SI{25}{\meter} above the ground plane.

\begin{figure}[t]
    \centering
    \includegraphics[width=\linewidth]{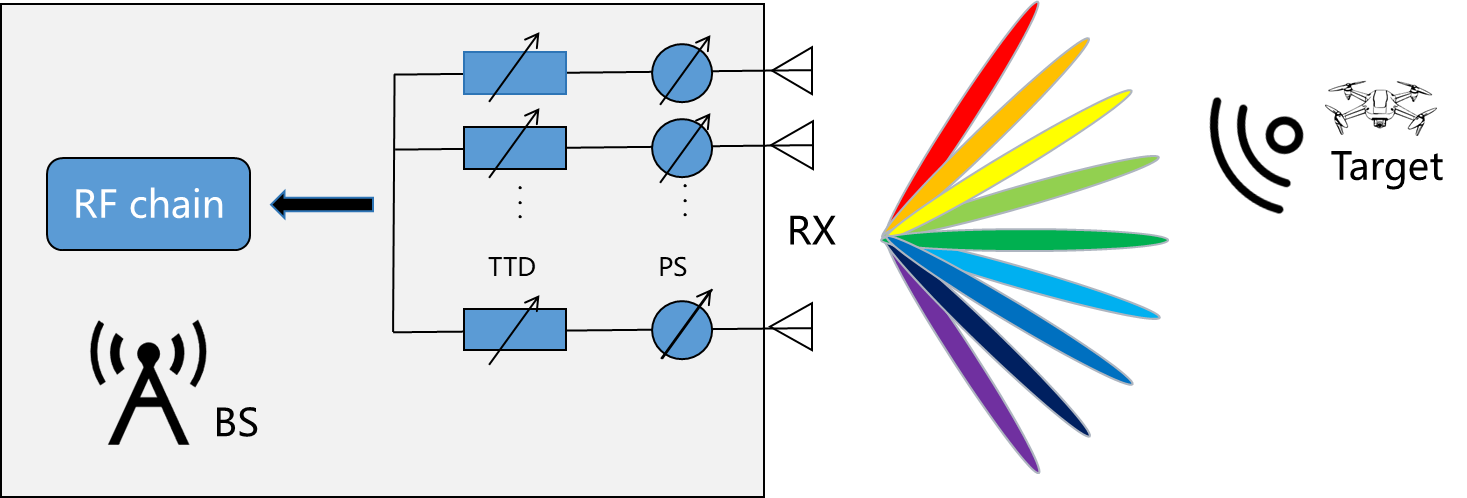}
    \caption{System architecture for uplink UAV localization with a phase-time array at the BS.}
    \label{fig:system_arch}
    \vspace{-5mm}
\end{figure}

\subsection{Rainbow Beamforming}
In wideband mmWave systems, the beam squint effect arises when a conventional, frequency-independent phase-shifter array is used. The physical propagation delay across the array aperture causes the beam direction to naturally shift with frequency. 
This is often seen as a performance-degrading impairment for communication, but can be deliberately controlled and exploited for sensing.

By augmenting the phase-shifter array with TTD elements, it becomes possible to synthesize a desired frequency-to-space mapping. The beamforming weight vector $\vect{w}_m \in \mathbb{C}^{N_r \times 1}$ for the $m$-th subcarrier is:
\begin{equation}\label{eq:beamforming_weight}
[\vect{w}_m]_n = \frac{1}{\sqrt{N_r}} \exp\left(-\j \left(\phi_n + 2\pi (f_m - f_0) \tau_n\right)\right),
\end{equation}
where $f_m$ is the frequency of the $m$-th subcarrier and $f_0$ is the carrier frequency, $\phi_n$ and $\tau_n$ are the phase shift and time delay applied at the $n$-th antenna. In our framework, the phase shift and time delay values are learnable parameters. They are initialized based on the strategy in \cite{luo2023beam} to span a predefined angular range but are subsequently optimized during training.

The received uplink signal $y_m$ at the BS on subcarrier $m$ after receive beamforming is expressed as:
\begin{equation}\label{eq:received_signal}
y_m = \sqrt{\frac{P_t}{M}} \vect{w}_m^H \vect{h}_m x_m + \eta_m,
\end{equation}
where $\vect{h}_m \in \mathbb{C}^{N_r \times 1}$ is the channel vector, $P_t$ is the transmit power of UAV, $x_m$ is the transmitted symbol, and $\eta_m$ is i.i.d. complex additive white Gaussian noise (AWGN). The collection of received samples, $\vect{y} = [y_1, \dots, y_{M}]^T \in \mathbb{C}^{M \times 1}$, forms the observation vector.


\begin{figure}[t]
    \centering
    \begin{subfigure}[b]{0.32\linewidth}
        \centering

        \includegraphics[height=1.5cm]{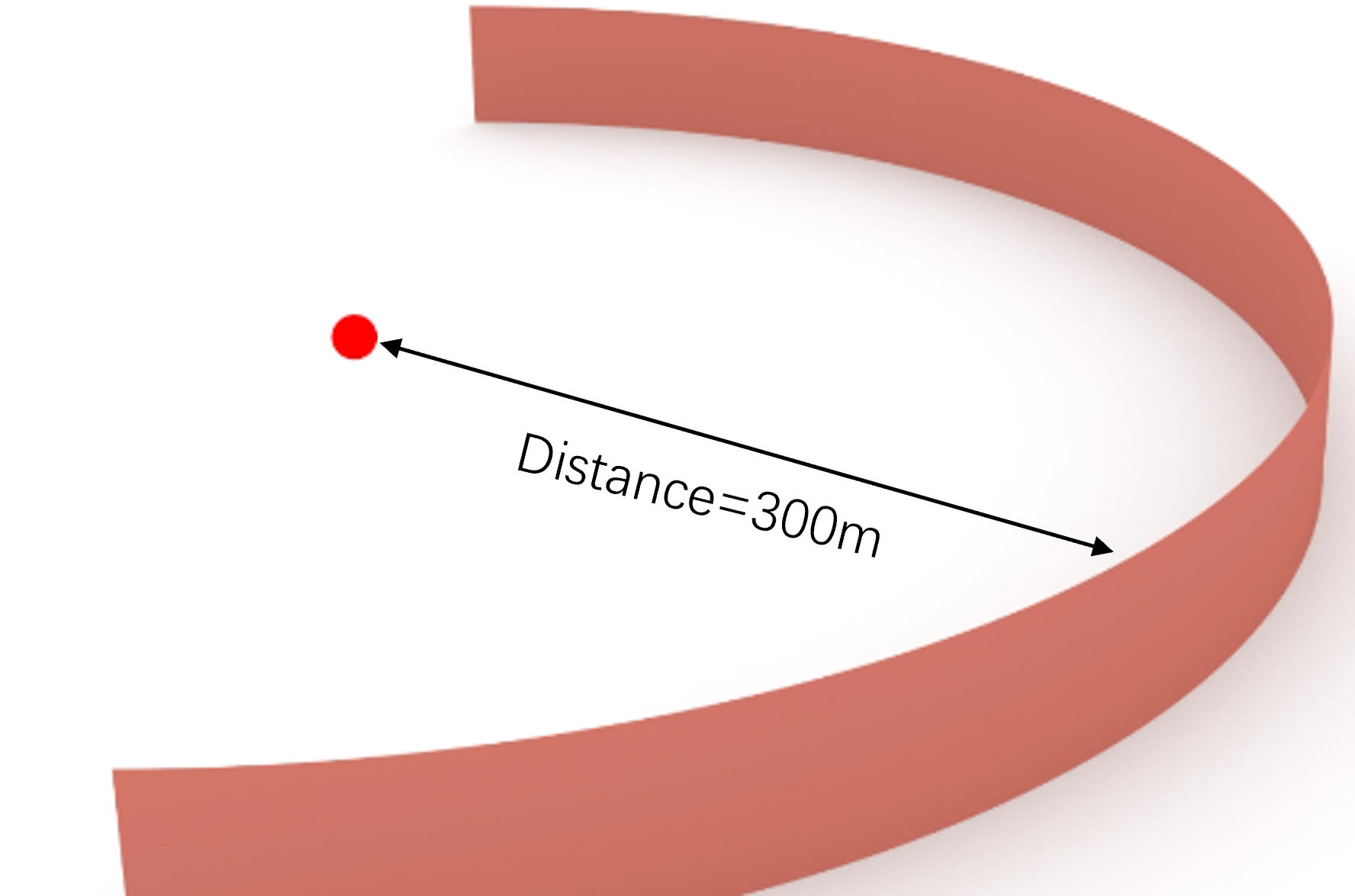} 
        \caption{`circle' scene}
        \label{fig:circle}
    \end{subfigure}
    \hfill
    \begin{subfigure}[b]{0.32\linewidth}
        \centering

        \includegraphics[height=1.5cm]{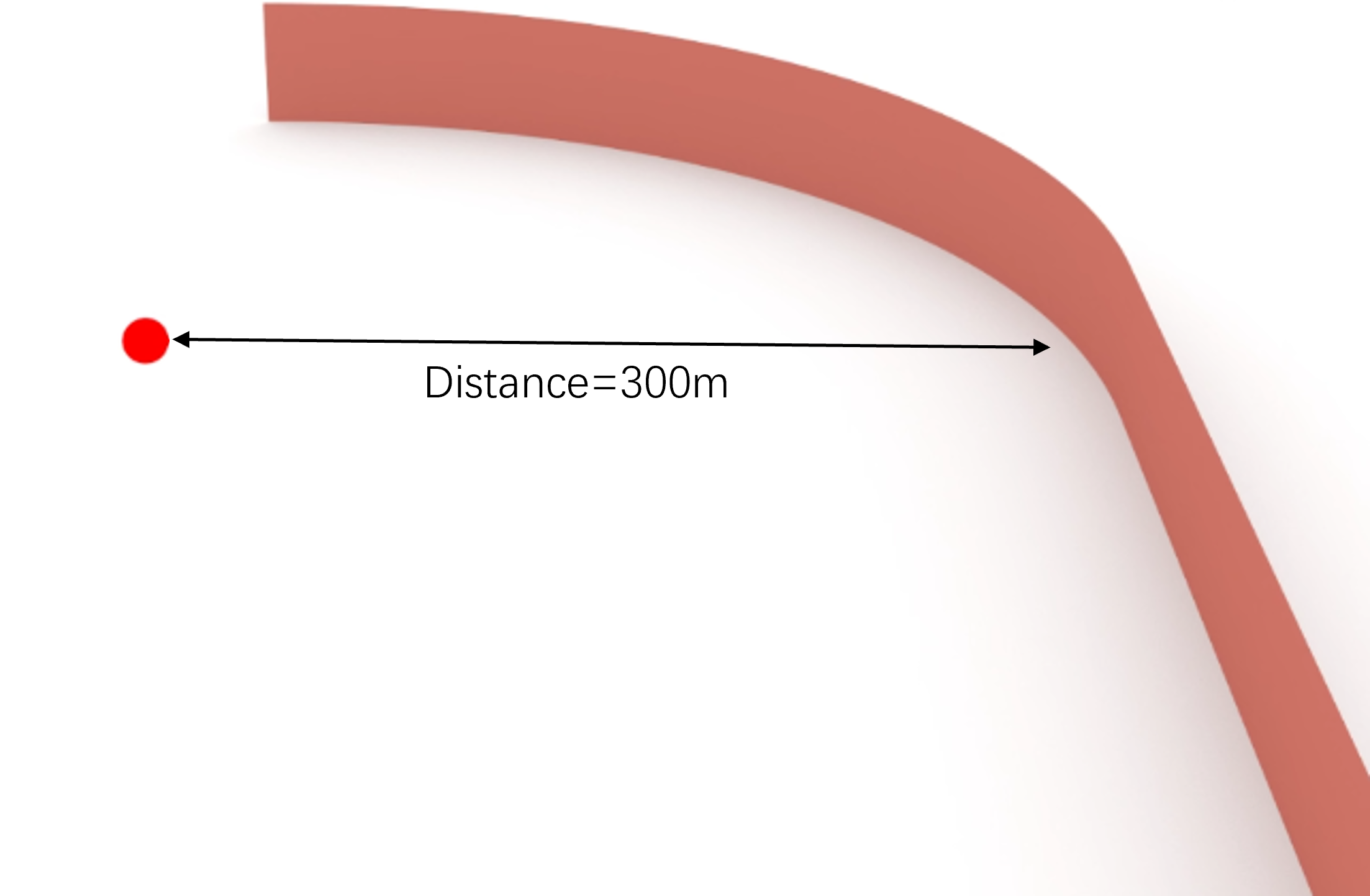}
        \caption{`rounded-L' scene}
        \label{fig:irregular}
    \end{subfigure}
    \hfill
    \begin{subfigure}[b]{0.32\linewidth}
        \centering

        \includegraphics[height=1.5cm]{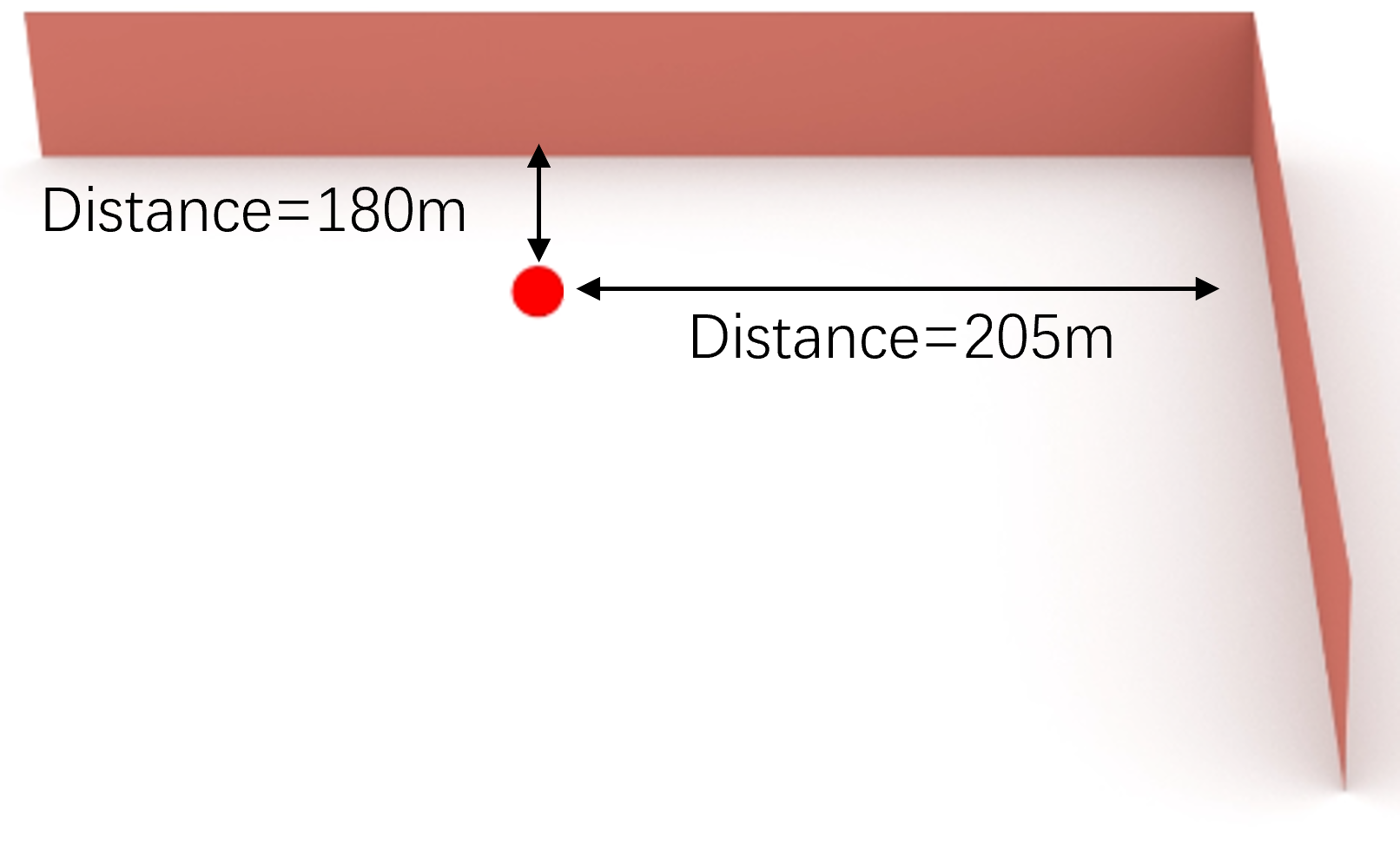}
        \caption{`L' scene}
        \label{fig:L}
    \end{subfigure}
    
    \caption{Three multipath scenes modeled by Blender.}
    \label{fig:scenes}
\end{figure}
\subsection{High-Fidelity Ray-Tracing Channel Generation}
To ensure our model is trained on physically realistic channel data, we employ a meticulous simulation pipeline based on the Sionna ray-tracing library. We define four distinct 3D environments to evaluate performance under varying degrees of multipath propagation:
\begin{enumerate}
\item \textbf{LoS-only Scene:} \new{An unobstructed environment dominated by the direct LoS path, serving as a baseline.}
    
    \item \textbf{`circle' Scene:} \new{Features a semicircular wall (Fig.~\ref{fig:circle}) that confines reflected paths to a narrow azimuth range, generating \textbf{uni-directional multipath with low angular spread}.}
    
    \item \textbf{`rounded-L' Scene:} \new{Composed of a quarter-circle arc and a straight wall (Fig.~\ref{fig:irregular}). This structure introduces diverse arrival directions, generating \textbf{multipath with high angular spread}.}
    
    \item \textbf{`L' scene:} \new{Consists of two perpendicular walls representing an urban street corner. This geometry generates a \textbf{rich multipath profile including higher-order reflections} (e.g., double-bounce paths).}

\end{enumerate}

For all scenes, the radio material properties are assigned to be concrete for the ground and brick for the walls. The Blender scene is imported into Sionna, and for each of the \num{100000} random UAV positions, Sionna's path solver traces interactions to compute all valid propagation paths and synthesize the final frequency-domain channel matrix $\mat{H}$.

\section{Proposed Localization Framework}
\begin{figure}[t]
    \centering
    \includegraphics[width=\linewidth]{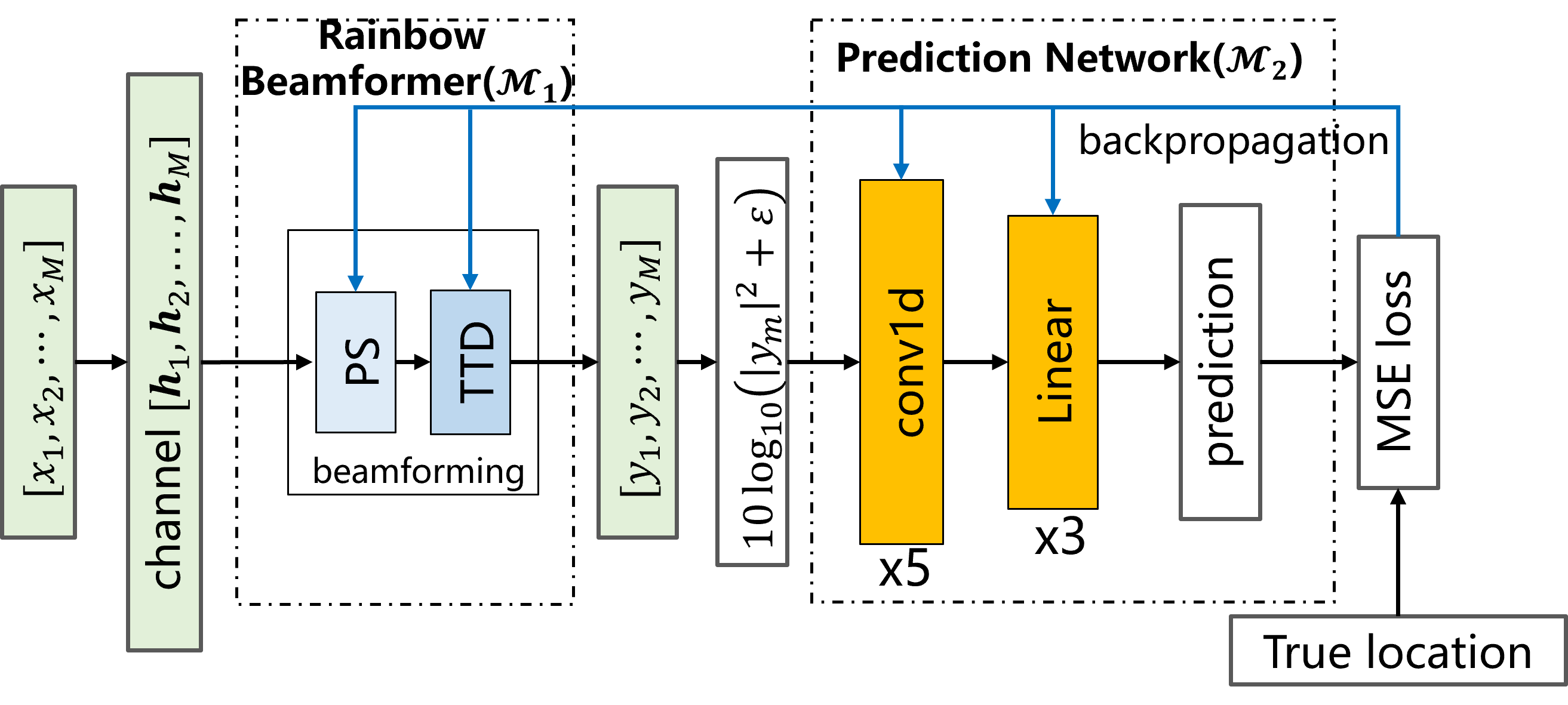} 
    \caption{Data flow of the MARBLE-Net architecture. }
    \label{fig:framework_arch}
    \vspace{-5mm}
\end{figure}

We introduce MARBLE-Net, a fully differentiable, end-to-end architecture that maps the raw channel matrix to a 2D coordinate estimate. It comprises two cascaded modules: a learnable rainbow beamformer ($\mathcal{M}_1$) and a prediction network ($\mathcal{M}_2$). As illustrated in Fig.~\ref{fig:framework_arch}, the channel data are first processed by $\mathcal{M}_1$ to generate a power spectrum, which is subsequently fed into $\mathcal{M}_2$ for localization.
The prediction error is then backpropagated through both modules, enabling joint optimization of the physical-layer beamforming and the prediction network parameters.


\subsection{Learnable Rainbow Beamformer ($\mathcal{M}_1$)}
This module acts as the physical-layer front end. Its trainable parameters consist solely of the PS and TTD vectors, $\bm{\phi}, \bm{\tau} \in \mathbb{R}^{N_r}$.
Given the channel matrix, it computes the received spectrum $\vect{y}$ using \eqref{eq:beamforming_weight} and \eqref{eq:received_signal}, and produces the log-magnitude power spectrum
$\mathbf{P}_{\text{dB}} \in \mathbb{R}^{M}$ with each element defined as $\mathbf{P}_{\text{dB},m} = 10 \log_{10}(|y_m|^2 + \epsilon)$.

\subsection{Prediction Network ($\mathcal{M}_2$)}
The prediction network is a 1D CNN designed to regress the target's 2D coordinates from the 1D power spectrum.
\begin{itemize}
    \item \textbf{Feature Extractor:} A stack of five \texttt{Conv1d} layers with progressive downsampling extracts hierarchical spectral features, effectively capturing multi-scale patterns induced by multipath interference.
    \item \textbf{Regression Head:} An MLP with three hidden layers processes the flattened feature vector. The final output layer has 2 neurons for the $(x, y)$ coordinates.
\end{itemize}

The output is rescaled by the maximum possible distance to yield the final estimate. The entire MARBLE-Net model is trained to minimize the MSE loss.

\subsection{Multi-Stage Training Strategy}
\new{To address the non-convex loss landscape and ensure stable convergence, we propose a multi-stage strategy transitioning from physical-layer initialization to end-to-end optimization.}

\begin{itemize}
    \item \textbf{Stage 1: Power-Based Beamformer Initialization.} 
    \new{We optimize the PS and TTD parameters of $\mathcal{M}_1$ to maximize the \textbf{total received signal power} $\sum_m {\mathbf{P}_{\text{dB,m}}}$, while keeping $\mathcal{M}_2$ fixed. This guarantees a high-SNR input, preventing the network from fitting to noise during the initial phase.}
    
    \item \textbf{Stage 2: Alternating Localization Optimization.} 
    \new{We iteratively update the prediction network $\mathcal{M}_2$ and the beamformer $\mathcal{M}_1$ in an alternating manner. Crucially, the beamformer is now updated to minimize the \textbf{localization error} rather than maximizing power, implicitly steering the beam pattern to exploit geometrically informative multipath components.}

    \item \textbf{Stage 3: End-to-End Joint Fine-tuning.} 
    \new{Finally, all parameters are unfrozen to jointly train the entire MARBLE-Net architecture. }
\end{itemize}


\section{Simulation Results}
\subsection{Simulation Setup}
We generate a dataset of \num{100000} samples for each of the four scenes using the Sionna pipeline, with key parameters listed in Table~\ref{tab:sim_params}. Each dataset is split into training (80\%), validation (10\%), and testing (10\%) sets. All models are implemented in PyTorch and trained using the Adam optimizer. MARBLE-Net is trained using the multi-stage strategy described in Section III-C.

\begin{table}[t]
\caption{Key Simulation Parameters}
\label{tab:sim_params}
\centering
\begin{tabular}{l l l}
\toprule
\textbf{Parameter} & \textbf{Symbol} & \textbf{Value} \\
\midrule
Start Frequency & $f_0$ & \SI{28}{\giga\hertz} \\
Bandwidth & $W$ & $\approx \SI{380}{\mega\hertz}$ \\
Subcarrier Spacing & $\Delta f$ & \SI{240}{\kilo\hertz} \\
Number of Subcarriers & $M$ & 1584 \\
BS Antennas (ULA) & $N_r$ & 128 \\
UAV Antenna & $N_t$ & 1 \\
UAV Transmit Power & $P_t$ & 23, 18, 13 {dBm} \\
BS/UAV Height & - & \SI{25}{\meter} \\
UAV-BS Distance Range & - & [\SI{5}{m}, \SI{200}{m}] \\
UAV Azimuth Range & - & [$-60\degree$, $60\degree$] \\
Ray-Tracing Max Depth & - & 2 \\
Dataset Size & - & \num{100000} samples \\
\bottomrule
\vspace{-5mm}
\end{tabular}
\end{table}

\begin{table*}[!t]

\caption{Performance comparison and ablation study of different localization methods across the four scenes}
\label{tab:results_comparison}
\centering
\begin{tabular}{@{}llccccc@{}}
\toprule
\multirow{2}{*}{\textbf{Scene}} & \multirow{2}{*}{\textbf{Metric}} & \textbf{k-NN} & \textbf{RaiNet \cite{klus2025deep}} & \textbf{RaiNet} & \textbf{MARBLE-Net} & \textbf{MARBLE-Net} \\
& & \textbf{($K=5$)} & \textbf{(fixed)} & \textbf{(adaptive)} & \textbf{(fixed)} & \textbf{(adaptive)} \\
\midrule
\multirow{3}{*}{LoS-only} 
& Localization RMSE (m) & 15.73 & 3.72 & 1.96 & 1.70 & \textbf{1.52} \\
& Angle RMSE (\degree)  & 0.13  & 1.18 & 1.26 & 1.28 & \textbf{2.42} \\
& Range RMSE (m)        & 15.73 & 3.63 & 1.89 & 1.66 & \textbf{1.44} \\
\midrule
\multirow{3}{*}{`circle'} 
& Localization RMSE (m) & 16.63 & 4.12 & 2.85 & 2.84 & \textbf{2.30} \\
& Angle RMSE (\degree)  & 0.18  & 1.22 & 1.14 & 2.55 & \textbf{1.39} \\
& Range RMSE (m)        & 16.63 & 4.01 & 2.76 & 2.79 & \textbf{2.26} \\
\midrule
\multirow{3}{*}{`rounded-L'} 
& Localization RMSE (m) & 13.32 & 3.76 & 2.55 & 2.24 & \textbf{2.15} \\
& Angle RMSE (\degree)  & 0.17  & 1.21 & 1.12 & 0.72 & \textbf{0.70} \\
& Range RMSE (m)        & 13.32 & 3.56 & 2.27 & 2.14 & \textbf{2.04} \\
\midrule
\multirow{3}{*}{`L'} 
& Localization RMSE (m) & 5.85  & 1.33 & 0.86 & 0.79 & \textbf{0.61} \\
& Angle RMSE (\degree)  & 0.14  & 0.62 & 0.41 & 0.25 & \textbf{0.17} \\
& Range RMSE (m)        & 5.84  & 1.21 & 0.78 & 0.74 & \textbf{0.58} \\
\bottomrule
\end{tabular}
\end{table*}
\begin{figure*}[t!]
    \centering
    \begin{subfigure}[b]{0.22\linewidth}
        \centering
        \includegraphics[height=6cm]{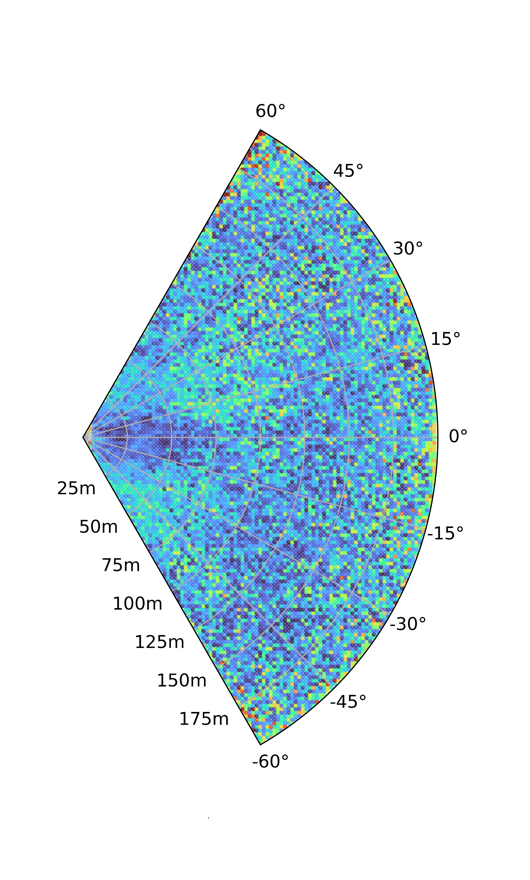}
        \caption{LoS-only Scene}
        \label{fig:error_los}
    \end{subfigure}
    \hfill
    \begin{subfigure}[b]{0.22\linewidth}
        \centering
        \includegraphics[height=6cm]{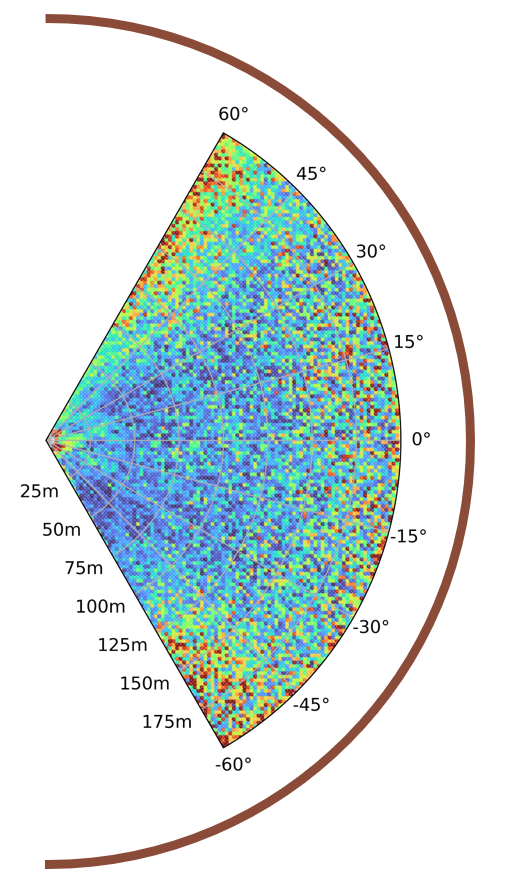}
        \caption{`circle' Scene}
        \label{fig:error_circle}
    \end{subfigure}
    \hfill
    \begin{subfigure}[b]{0.22\linewidth}
        \centering
        \includegraphics[height=6cm]{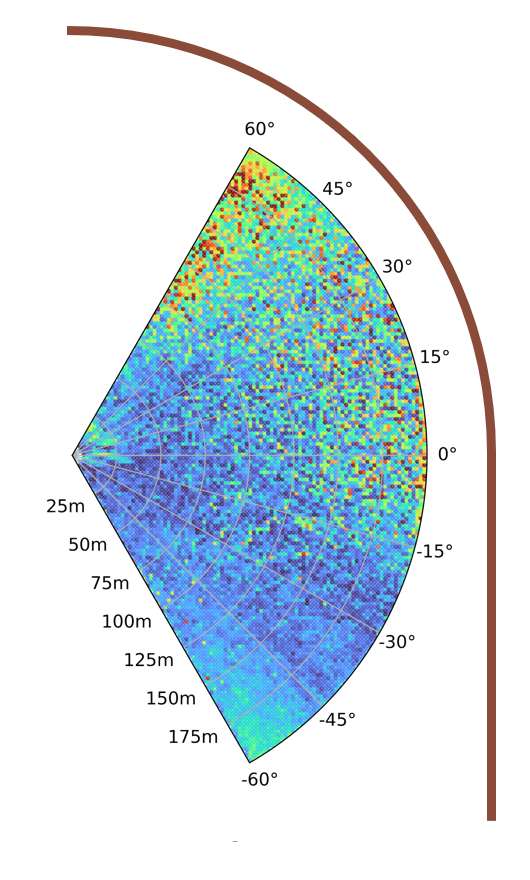}
        \caption{`rounded-L' Scene}
        \label{fig:error_irregular}
    \end{subfigure}
    \hfill
    \begin{subfigure}[b]{0.22\linewidth}
        \centering
        \includegraphics[height=6cm]{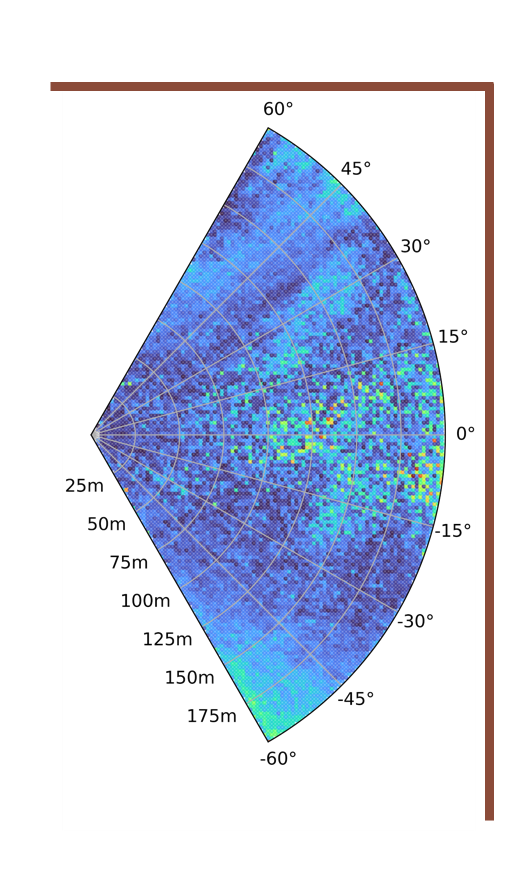}
        \caption{`L' Scene} 
        \label{fig:error_L}
    \end{subfigure}
    \hfill
    \begin{subfigure}[b]{0.08\linewidth}
        \centering
        \includegraphics[height=6cm]{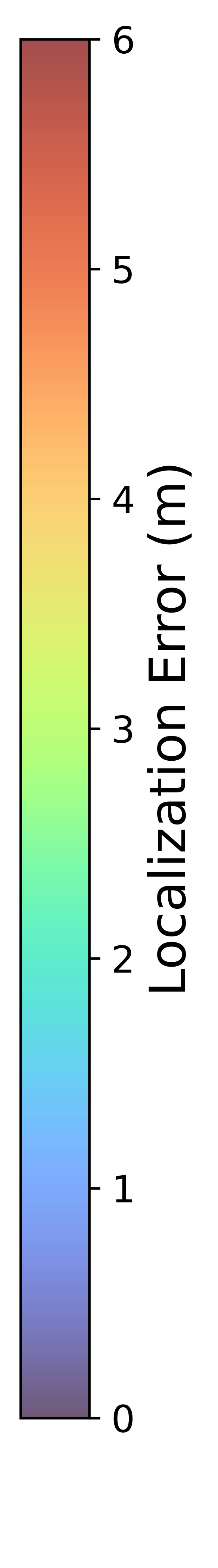}
    \end{subfigure}

    \caption{Spatial distribution of the localization error of the MARBLE-Net (adaptive) at $P_t=\SI{23}{dBm}$ across the four scenes.} 
    \label{fig:error_distribution}
\end{figure*}

\subsection{Baseline}

\subsubsection{k-Nearest Neighbors (k-NN)}
To benchmark against a traditional, non-learning-based method, we implement a k-NN algorithm with $K=5$. A codebook is constructed by storing the received power spectra (from a fixed initial beam) for 80,000 training samples (codewords) and their corresponding locations. At inference time, the spectrum from a test target is compared against all codebook entries using Euclidean distance. The coordinates of the $K$ nearest neighbors are averaged to produce the location estimate.

\subsubsection{RaiNet}

As a point of comparison, the baseline RaiNet given in \cite{klus2025deep} adopted a fixed rainbow beam and only trained the prediction network. This represents a conventional DL-based approach with a non-adaptive sensing beam. The RaiNet architecture consists of a feature extractor and a classifier. The feature extractor contains three 1D convolutional layers, followed by a classifier with three fully connected layers. The \texttt{Tanh} function is used as the primary activation function throughout the network.
\subsection{Performance Comparison and Ablation Study}
The localization performance of the proposed MARBLE-Net, alongside several baselines, is comprehensively evaluated across the four distinct environments. The results, summarized in Table~\ref{tab:results_comparison}, provide a detailed ablation study on the impact of multipath complexity, joint beamformer-network training, and network architecture.

First, it is evident that all deep learning approaches outperform the traditional k-NN baseline. In the `circle' scene, for instance, k-NN yields a large localization RMSE of \SI{16.63}{m}, underscoring its inability to handle the complex signal fingerprints created by multipath propagation. Template-matching methods like k-NN fail when the direct relationship between the signal spectrum and geometric location is severely distorted, rendering the pre-compiled codebook ineffective.

To deconstruct the sources of performance gain, we compare four DL-based configurations: RaiNet with a fixed beamformer (`RaiNet (fixed)'), RaiNet with joint training where the beamformer is adaptive to the scene (`RaiNet (adaptive)'), MARBLE-Net's architecture with a fixed beamformer (`MARBLE-Net (fixed)'), and a scene-adaptive beamformer  (`MARBLE-Net (adaptive)'). This allows us to isolate the effects of network architecture and joint training.

\textbf{The Benefit of Joint Training:}
Table~\ref{tab:results_comparison} highlights the critical importance of joint end-to-end training between the beamformer and the localization network. Comparing the `fixed' and `adaptive' configurations reveals consistent performance improvements across all scenes. For MARBLE-Net, joint training reduces the localization RMSE by 11\%, 19\%, 4\%, and 23\% in the LoS-only, `circle', `rounded-L', and `L' scenes, respectively, while RaiNet achieves even larger relative reductions of 47\%, 31\%, 32\%, and 35\%. These results demonstrate that enabling beamformer adaptation with respect to the environment is essential for effectively exploiting multipath information.

The relative gain from joint training differs across architectures. The simpler RaiNet benefits more significantly from beamformer adaptation due to its limited representational capacity and higher sensitivity to the quality of physical-layer observations. In contrast, MARBLE-Net already attains strong performance with a fixed beamformer, resulting in smaller but consistent additional gains from joint optimization. This confirms that network capacity and adaptive beamforming play complementary roles in aligning the sensing strategy with the underlying geometry of multipath propagation.

\textbf{The Nuanced Impact of Multipath:}
Table~\ref{tab:results_comparison} further reveals a non-monotonic relationship between multipath propagation and localization accuracy. Compared to the LoS-only case, performance degrades in the highly complex `circle' scene, where dense and uni-directional multipath introduces severe geometric ambiguities. For MARBLE-Net, the localization RMSE increases from \SI{1.52}{m} to \SI{2.30}{m}, indicating that such confined multipath remains challenging to resolve even with adaptive beamforming.

In contrast, environments with more structured and directionally diverse multipath exhibit improved localization performance. In the `rounded-L' scene, directional NLoS components provide additional spatial diversity, while in the `L' scene, two perpendicular planar walls generate stable and geometrically informative specular reflections. Consequently, the localization RMSE of MARBLE-Net is reduced to \SI{0.61}{m}, outperforming all other scenarios, including the LoS-only case. This demonstrates that structured NLoS paths can be effectively exploited as \emph{virtual anchors} through joint optimization of sensing beam and the location network.

\textbf{The Benefit of Network Complexity:}
Table~\ref{tab:results_comparison} also shows that increasing network capacity yields clear performance gains. With a fixed beamformer, the more complex MARBLE-Net architecture (`MARBLE-Net (fixed)') consistently outperforms `RaiNet (fixed)' across all scenes, for example reducing the localization RMSE from \SI{3.72}{m} to \SI{1.70}{m} in the LoS-only scene. This reflects the stronger capability of a high-capacity network to model the complex signal–location relationships induced by multipath propagation.

Nevertheless, network complexity alone is insufficient to fully exploit multipath-rich environments. The most significant performance improvements are achieved when a high-capacity prediction network is jointly optimized with a learnable, scene-adaptive beamformer, as evidenced by the superiority of `MARBLE-Net (adaptive)'. 

\begin{figure}[t]
    \centering
    \setlength{\abovecaptionskip}{3pt}
    \setlength{\belowcaptionskip}{-8pt}
    \includegraphics[width=0.9\linewidth]{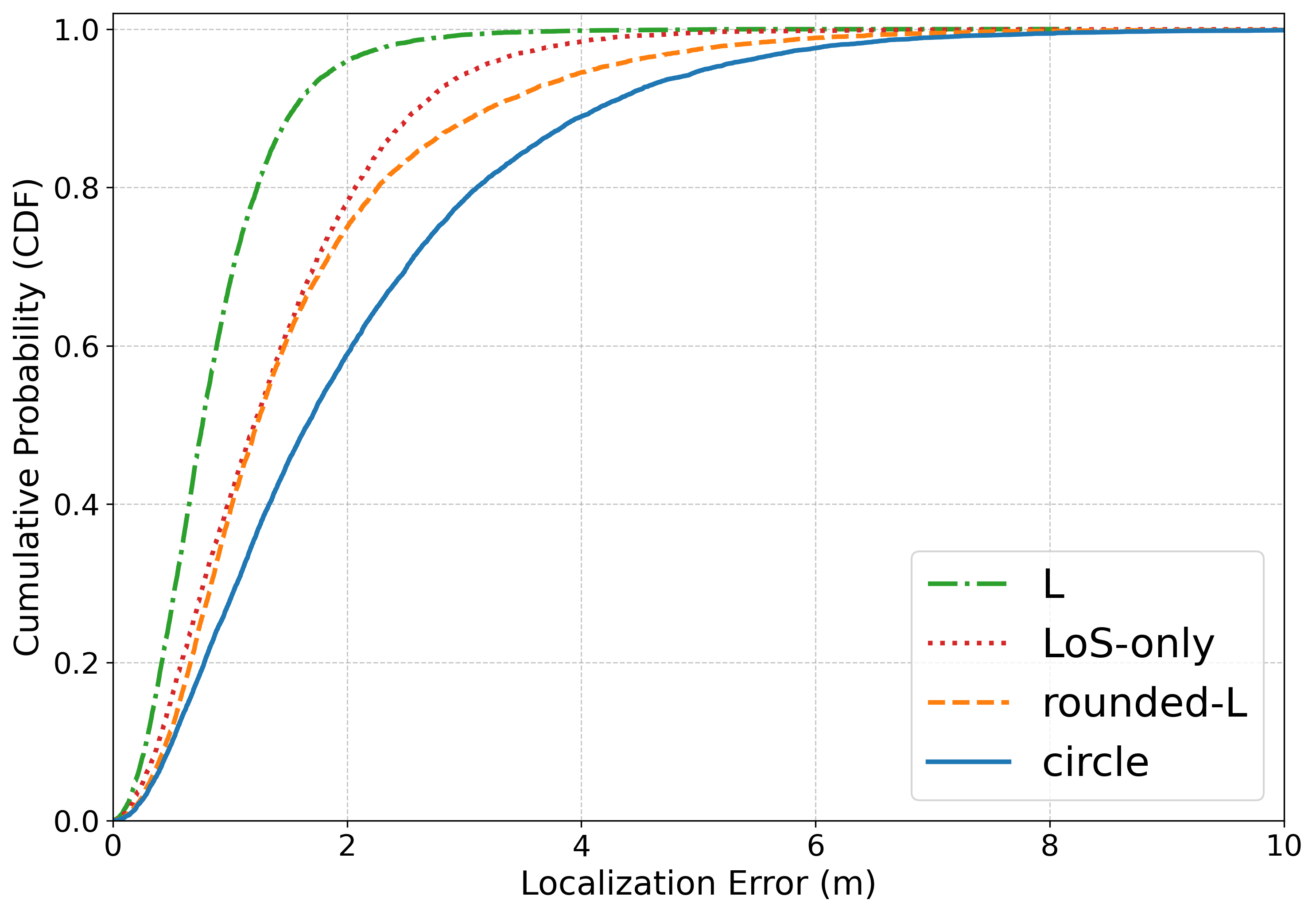}
    \caption{\small Localization error CDF of MARBLE-Net at $P_t = 23$ dBm.}
    \label{fig:loc_cdf_23dbm}
\end{figure}

\begin{figure}[t]
    \centering
    \setlength{\abovecaptionskip}{3pt}
    \setlength{\belowcaptionskip}{-8pt}
    \includegraphics[width=0.9\linewidth]{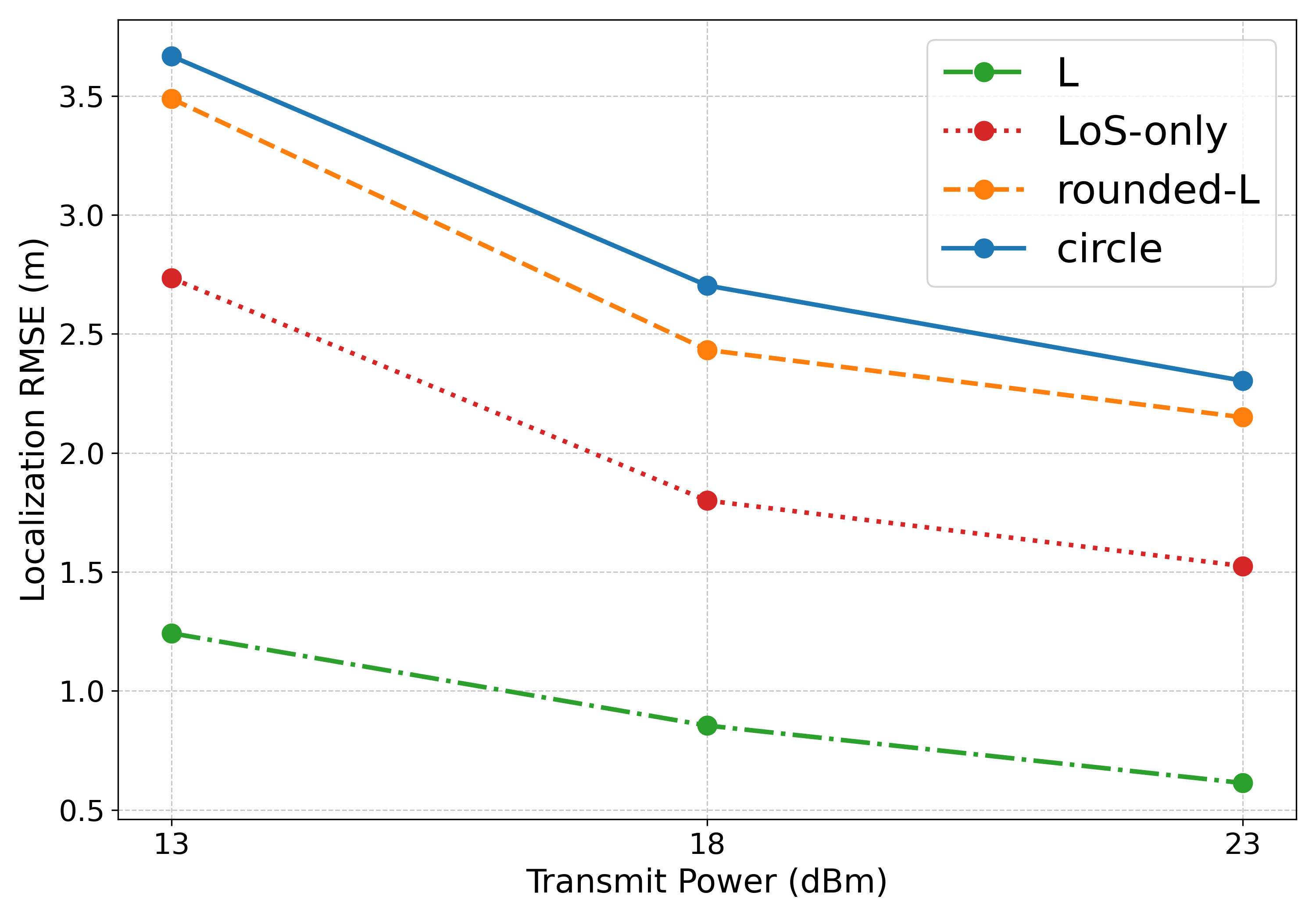}
    \caption{\small Localization RMSE of MARBLE-Net vs. $P_t$.}
    \label{fig:loc_rmse_vs_pt}
\end{figure}

\subsection{Qualitative and Error Distribution Analysis}

To gain deeper insight into the system's behavior, Fig.~\ref{fig:error_distribution} visualizes the spatial error distribution of the MARBLE-Net (adaptive) model, which is consistent with the quantitative results in Table~\ref{tab:results_comparison}. 
The `L' scene (Fig.~\ref{fig:error_distribution}d) exhibits the best performance, with most regions characterized by low localization errors. 
The `rounded-L' scene (Fig.~\ref{fig:error_distribution}c) shows strong overall performance but reveals spatial heterogeneity: the upper region near the semi-circular arc suffers from larger errors due to curved geometry inducing uni-directional reflections with limited angular spread, whereas the lower region near the straight wall achieves higher precision, as planar surfaces provide structured and directional specular reflections that can be effectively exploited as virtual anchors. 
This comparison highlights the nuanced role of multipath propagation: structured, directional multipath from planar geometries serves as a valuable localization resource, potentially outperforming LoS-only conditions, while confined uni-directional multipath from curved structures introduces ambiguities that degrade performance.

The cumulative distribution function (CDF) of localization error in Fig.~\ref{fig:loc_cdf_23dbm} provides further insight into performance consistency across different environments. The `L' scene exhibits the most favorable distribution, with a steep rise reaching nearly 100\% probability at a \SI{2}{m} error threshold, indicating high localization stability. The LoS-only case follows as the second-best scenario, while the `rounded-L' and `circle' scenes show more gradual curves, with the `rounded-L' performance lying between the two. The `circle' scene exhibits the least favorable distribution, reflecting the difficulty of resolving signal ambiguities in uni-directional multipath environments. These trends are consistent with the results in Table~\ref{tab:results_comparison}, confirming that structured multipath is highly beneficial for localization, whereas the geometry of the `circle' scene remains a major limitation.

\subsection{Impact of Transmit Power}

We evaluate the system's robustness to noise by varying the UAV transmit power, as shown in Fig.~\ref{fig:loc_rmse_vs_pt}. As expected, the localization RMSE improves for all scenes with increased transmit power due to the higher SNR. Notably, the performance hierarchy across scenes remains consistent, with the `L' scene consistently outperforming the LoS-only case, while the `rounded-L' and `circle' scenes exhibit higher errors due to their respective multipath complexities. This confirms that the geometric impact of multipath (whether beneficial or detrimental) is consistent across different noise levels.

\subsection{Complexity and Latency Analysis}
 
\begin{table}[t]
\caption{Complexity and latency comparison}
\label{tab:latency_comparison}
\centering
\begin{tabular}{@{}lcc@{}}
\toprule
\textbf{Method} & \textbf{Parameters / Memory} & \textbf{Avg. Latency/Sample} \\
\midrule
MARBLE-Net & 13.96 M params & \textbf{\SI{0.0172}{ms}} \\
RaiNet \cite{klus2025deep} & 1.95 M params & \SI{0.0180}{ms} \\
k-NN (GPU) & 484 MB codebook & \SI{2.2122}{ms} \\
\bottomrule
\vspace{-5mm}
\end{tabular}
\end{table}
Table \ref{tab:latency_comparison} compares computational complexity and latency, measured on an NVIDIA RTX 5880 GPU with dual AMD EPYC 7Y43 CPUs.
\new{Although MARBLE-Net has significantly more parameters than RaiNet (13.96 M vs. 1.95 M), its average inference latency (\SI{0.0172}{ms}) is marginally lower (\SI{0.0180}{ms}). This result is attributed to the GPU's ability to efficiently parallelize the wider layers in MARBLE-Net, combined with the use of the efficient LeakyReLU activation, as opposed to the more expensive Tanh function used in RaiNet.}
Both models achieve over two orders of magnitude faster inference than the k-NN baseline (\SI{2.21}{ms}), which depends on a 484 MB codebook and costly neighbor search.
These results confirm the proposed model's balance between accuracy and efficiency, enabling real-time deployment.




\section{Conclusion}
This paper has introduced MARBLE-Net, an end-to-end deep learning framework that jointly optimizes an adaptive rainbow beamformer and a localization network for ISAC systems.
By leveraging ray-tracing-generated datasets across environments of varying complexity, we have demonstrated that a data-driven approach can improve localization performance. Our results show that co-designing the sensing beam with the neural network processor, guided by a structured multi-stage training strategy, allows the system to learn to exploit multipath information, reducing localization error from \SI{5.85}{\meter} and \SI{1.33}{\meter} to \SI{0.61}{\meter} in a multipath-rich scene. This work validates that for high-precision localization in complex environments, the joint design of the physical-layer sensing strategy and the higher-layer processing network is not merely beneficial, but essential. Our findings also highlight a nuanced interaction with the propagation environment: while confined uni-directional multipath can hinder performance, we show that highly structured and directional multipath, as observed in the `L' scene, can be actively exploited to achieve accuracy surpassing even LoS-only conditions, turning multipath from an impairment into a localization resource.

Future work will extend the framework to 3D multi-target localization and adaptation in dynamic environments.

\bibliographystyle{IEEEtran}
\bibliography{reference} 

\end{document}